# Fretiq: Browser-Native Electric Guitar String Classification via Engineered Spectral Features and Held-Out Free-Play Evaluation

*Aadi Garg*
Department of Physics, California Polytechnic State University, San Luis Obispo
agarg35@calpoly.edu

**Abstract**

Identifying which string produces a given pitch in monophonic electric guitar audio is a fundamental classification challenge: a single pitch can often be produced on multiple strings at different fret positions, with timbral differences that prior listening studies confirm are largely imperceptible to untrained humans. Existing approaches using support vector machines and spectral envelope features have achieved F-measures of 0.90 for six-string electric guitar classification, while String-Inverse Frequency features in earlier work achieved F1 scores up to 0.72. We present Fretiq, a preliminary single-instrument, single-player browser-based string classification system built on a 26-dimensional feature representation integrating frequency band energies, spectral statistics, and 13 Mel-Frequency Cepstral Coefficients. Across five random seeds, a shuffled frame-level validation split yields 97.25% ± 0.32 accuracy on the full model, and an ablation study identifies MFCCs as the primary accuracy driver (92.09% ± 0.50 without MFCCs). We additionally introduce Comparison Training -- a data collection methodology in which canonical same-pitch pairs on adjacent strings are recorded in deliberate alternation. An initial evaluation using independently shuffled train/validation splits for each training condition found no net accuracy benefit; however, because conditions with different training-set compositions produce different, non-comparable shuffled validation sets, this comparison was confounded and is superseded here. Using a matched evaluation -- both conditions evaluated on an identical recording session held out from training, early stopping, and model selection -- including the comparison-session data improves overall accuracy on the matched open-string evaluation session by 25.78 ± 1.46 percentage points across five seeds, positive for all five seeds. A size-matched control, in which the comparison condition is downsampled to the same training-set size as the baseline, reproduces a numerically similar effect (25.26 ± 1.51 points), showing that total training-set size alone does not explain the effect; the comparison-session data may still contribute through mechanisms other than the deliberate alternation of adjacent-string pairs specifically -- the remaining effect could reflect deliberate alternation, differences in pitch and fret-position coverage, attack or dynamic variation, or other session-specific characteristics. Because a shuffled frame-level split allows adjacent frames from the same sustained note to appear in both training and validation, we additionally report an evaluation that withholds an entire recording session -- training on two full recording sessions and evaluating on a third session held out from training and model selection -- which yields a substantially lower 86.53% ± 1.23 accuracy, numerically close to an independently collected held-out free-play evaluation (87.8% on 103,000 frames). This demonstrates that shuffled frame-level validation produces a substantially more optimistic estimate than either the recording-session-held-out evaluation or the separately collected free-play evaluation. We describe the feature extraction pipeline implemented in both Python and TypeScript to guarantee training-inference parity, and document two critical implementation failure modes encountered during development. The system runs entirely in-browser with no hexaphonic pickup, fretboard sensor, camera, or multi-microphone setup required.

## 1. Introduction

The electric guitar is among the most expressive and widely played instruments in contemporary music, yet automatic analysis at the level of individual strings remains an open problem. Untrained listeners perform near chance in prior experiments, so computational systems must rely on subtle spectral cues rather than pitch alone. Unlike piano, where each key produces a unique pitch with no positional ambiguity, the guitar fretboard is inherently redundant: a given pitch is often playable on multiple strings at different fret positions, each with subtly different timbral characteristics arising from string gauge, tension, and the relative position of the pickup along the vibrating string length.

Perceptual studies confirm this difficulty. Abeßer (2013; originally presented in 2012) reports human-listener F-scores of 0.26 and 0.16 for bass and electric guitar string identification respectively — near chance level — underscoring how subtle the inter-string timbral difference truly is. For computational systems, the challenge is compounded in real-time settings where decisions must be made frame-by-frame from a single monophonic stream with no video or sensor data. Prior machine learning approaches have achieved F-measures of up to 0.90 for six-string electric guitar using SVMs with spectral envelope features, F1 up to 0.72 using String-Inverse Frequency estimation, and real-time string and fret detection for music education software (Dittmar et al., 2013). Prior systems address string and fret estimation in offline, desktop, or non-browser settings, but do not target browser-native deployment from a single DI stream or evaluate held-out free-play performance under that deployment setting. Fretiq bridges this gap by operating entirely in-browser from a single USB-C direct input signal, requiring no hexaphonic pickup, fretboard sensor, camera, or multi-microphone setup.

The system is built on a 26-dimensional engineered feature representation combining frequency band energies, spectral statistics, and Mel-Frequency Cepstral Coefficients. A multi-seed ablation study identifies MFCCs as the primary contributor to classification accuracy. We additionally describe Comparison Training, a structured data collection methodology in which acoustically ambiguous adjacent string pairs are recorded in deliberate alternation. An evaluation using a matched recording session held out from training, early stopping, and model selection finds a large and consistent benefit -- a 25.78 ± 1.46 percentage point improvement across five seeds, positive in all five -- that a size-matched control shows is not explained by increased training-set size alone. This supersedes an earlier, confounded comparison (Section 6.2) that had compared training conditions on different, non-comparable shuffled validation sets and found no net benefit.

A held-out free-play evaluation on 103,000 frames unseen during training yields 87.8% accuracy. A recording-session-held-out evaluation -- training on two full sessions and evaluating on a third session held out from training and model selection -- yields a numerically similar figure (86.53% ± 1.23), well below the 97.25% ± 0.32 obtained from a shuffled frame-level split on the same data. The agreement between these two unseen-data evaluations shows that the shuffled split is substantially more optimistic. The present design cannot fully separate the effects of adjacent-frame leakage from ordinary session-to-session and recording-protocol variation. This gap motivates temporal sequence modeling as future work, which we discuss in Section 8.

Two critical engineering failure modes encountered during development are documented as a contribution to reproducibility.

We position our work against existing literature in Section 2, before detailing the web architecture and 26-dimensional feature engineering pipeline in Sections 3 and 4. The Comparison Training protocol is introduced in Section 5, followed by the multi-seed MFCC ablation, matched held-out Comparison Training analysis, recording-session-held-out evaluation, and free-play evaluation in Section 6. Finally, we document the system's runtime failure modes (Section 7) and core limitations (Section 8) before concluding.

## 2. Related Work

### *2.1 Guitar String and Fret Classification*

Abeßer (2013; originally presented in 2012) presented the most directly comparable isolated string-detection study, reporting F = 0.93 for bass guitar and F = 0.90 for six-string electric guitar using spectral-envelope features and SVM classification. The same work reports human-listener F-scores of 0.26 and 0.16 for bass and electric guitar respectively, establishing the perceptual difficulty of the task. Kehling et al. (2014) later incorporated string estimation into a broader automatic tablature-transcription pipeline for electric guitar recordings, combining onset detection, multipitch estimation, and string classification into an end-to-end system, reporting 82% string-number accuracy as one component of the pipeline.

Geib, Schmitt, and Schuller (2017) introduced String-Inverse Frequencies (SIFs), features derived from the unfretted portion of the string, reporting F1 scores up to 0.72 for guitar and bass string detection and comparing SIFs against an MFCC + SVM baseline. Hjerrild and Christensen (2019) proposed a maximum a posteriori classifier requiring training data from only one fret per string, achieving 1.5% average absolute classification error across electric and acoustic guitar — though the method relies on a physically parameterized inharmonicity model rather than learned timbral features, making its assumptions different from Fretiq's data-driven browser pipeline. Dittmar et al. (2013) presented a real-time string and fret detection system for music education software using multi-pitch detection, feature extraction, and classification from polyphonic guitar recordings. Unlike Dittmar et al.'s real-time education-oriented system, Fretiq focuses on browser-native deployment from a single direct-input stream and reports a held-out free-play evaluation for its prototype setting.

### *2.2 Real-Time Guitar Technique Classification*

Fiorini et al. (2025) introduced the EG-IPT dataset and ipt~ Max/MSP object for real-time electric guitar playing technique classification, using CNN-based models across multiple audio source configurations and pickup configurations. Fretiq differentiates by operating entirely within a web browser from a single USB-C DI signal.

Stefani and Turchet (2025) demonstrated a real-time playing technique recognition pipeline for smart electro-acoustic guitar on a Raspberry Pi 4, classifying notes into eight playing techniques in real time. Their work identifies the lack of systems combining deep learning with accessible real-time constraints as an open gap — precisely the gap Fretiq targets, for electric guitar in the browser.

### *2.3 Guitar Tablature Estimation*

TabCNN (Wiggins and Kim, 2019) estimates full guitar tablature from solo acoustic guitar audio using constant-Q spectrograms and a 2D CNN. Unlike earlier MIDI-generated tablature datasets, TabCNN trains and tests on GuitarSet microphone recordings of real acoustic guitar performances; GuitarSet uses hexaphonic pickup signals for ground-truth annotation, while TabCNN itself estimates tablature from the monophonic microphone signal. Fretiq addresses a narrower subproblem — per-note electric guitar string identification — in a browser-based real-time DI setting.

### *2.4 MFCC-Based Timbre Classification*

Mel-Frequency Cepstral Coefficients are well established for instrument and timbre classification. Logan (2000) demonstrated their effectiveness for audio similarity, and subsequent work in instrument identification confirmed their utility for capturing timbral envelope information beyond raw spectral features. Here, we implement a complete MFCC pipeline without external audio libraries, enabling identical deployment in both Python training and browser inference.

## 3. System Architecture

Fretiq is implemented as a Next.js 16 application using the App Router with TypeScript. Three primary layers handle audio capture, feature extraction and classification, and 3D fretboard visualization. In the tested configuration, the audio context and interface operated at 44.1 kHz.

### *3.1 Audio Capture*

Audio is captured via the Web Audio API using getUserMedia with echo cancellation, noise suppression, and automatic gain control disabled. The guitar connects through a Boss Katana Gen 3 amplifier acting as a USB-C audio interface at 44.1 kHz. The audio graph runs MediaStreamSource through an 8x GainNode into an AnalyserNode configured with

fftSize=2048 and smoothingTimeConstant=0.75, producing 1024 frequency bins per frame at approximately 21.5 Hz per bin.

### *3.2 Pitch Detection*

Pitch detection uses the Pitchy library implementing the McLeod Pitch Method (MPM), applied to floating-point time-domain waveform data. A sliding clarity threshold accommodates the reduced McLeod scores characteristic of high-frequency notes: 0.90 below 200 Hz, 0.86 from 200–500 Hz, 0.80 from 500–800 Hz, and 0.72 above 800 Hz. Frequencies outside 70–1200 Hz are discarded. A sustain gate holds the last detected note for 400 ms and clears immediately if RMS intensity falls below 0.02.

### *3.3 String Classification Pipeline*

The detected MIDI pitch constrains which strings are physically capable of producing it within frets 0–24. The 1024-bin FFT magnitude spectrum is transformed into a 26-dimensional feature vector and passed through the trained neural network. Softmax scores for non-candidate strings are zeroed. A pitch-based penalty additionally discounts implausible assignments: E2, A2, and D3 receive 0.3x confidence multipliers when the detected MIDI pitch exceeds their practical playing range. Remaining scores are renormalized and passed through a temporal stability gate requiring two consecutive agreeing frames before committing a prediction, and six consecutive disagreeing frames before switching. Feature extraction and neural network inference (26-feature extraction + dense network forward pass) average 2.07 ms per animation frame (p95: 4.40 ms) on an Intel Core Ultra 9 275HX with 32 GB RAM in Chrome, well within the 16 ms budget of a 60 fps animation frame.

### *3.4 Visualization*

The fretboard visualization uses React Three Fiber with InstancedMesh across 144 instances, corresponding to 6 strings × 24 fretted positions; open-string positions are handled separately. During validation, all valid positions for the detected pitch render with brightness proportional to model confidence, forming a probability heatmap. Camera sits at [0, 3.5, 6] with 55° FOV for a guitarist's-eye view down the neck.

## 4. Feature Extraction

Each FFT frame is transformed into a 26-dimensional feature vector, implemented identically in Python (training) and TypeScript (inference) via pre-computed matrix representations. All matrices are computed once at module import. Frequency bin boundaries assume a 44.1 kHz sample rate and 1024 positive-frequency bins, giving approximately 21.5 Hz per bin.

### *4.1 Frequency Band Energies*

Eight frequency band energies capture mean magnitude across standard psychoacoustic bands:

| Band | Bin Range | Frequency Range (44.1 kHz) |
|---|---|---|
| Sub-bass | 1–5 | ~21–107 Hz |
| Bass | 5–20 | ~107–430 Hz |
| Low-mid | 20–60 | ~430–1290 Hz |
| Mid | 60–120 | ~1290–2580 Hz |
| Upper-mid | 120–200 | ~2580–4300 Hz |
| Presence | 200–350 | ~4300–7525 Hz |
| Brilliance | 350–600 | ~7525–12900 Hz |
| Air | 600–1024 | ~12900–22050 Hz |

*Table 1. Frequency band definitions computed at 44.1 kHz, 1024 bins.*

### *4.2 Spectral Statistics*

Five spectral statistics round out the non-MFCC features: spectral centroid (frequency-weighted mean of the magnitude spectrum), spectral rolloff (bin below which 85% of energy is concentrated), spectral flatness (ratio of geometric to arithmetic mean, measuring tonal vs. noise-like character), peak bin index, and peak magnitude. Combined with the eight band energies, this yields 13 non-MFCC features.

### *4.3 Mel-Frequency Cepstral Coefficients*

Thirteen MFCCs are computed without external audio libraries. A 40-filter mel-spaced filterbank is pre-computed as a 40×1024 matrix using triangular filters. Log compression applies a floor of 1e-6. An orthonormal DCT-II matrix of shape 13×40 is pre-computed; MFCCs are the matrix product of the DCT matrix with the log mel-filtered spectrum, normalized by dividing by 20 and clipping to [−1, 1]. Per-frame extraction reduces to two matrix multiplications plus element-wise operations — fast enough for the browser's animation frame budget.

## 5. Comparison Training

### *5.1 The String Ambiguity Problem*

Same-pitch notes on adjacent strings are acoustically similar but not identical. Several physical factors may contribute to the difference: string gauge (thicker strings have more inharmonicity and stronger low-order partials), string tension at the fret position (affecting sustain and brightness), and the distance from the pickup to the vibrating node. As confirmed by Abeßer (2013; originally presented in 2012), these differences frequently fall below the threshold of conscious human discrimination.

Standard data collection — recording each string in sequence across the neck — gives the model positive examples per class but no explicit exposure to the hardest confusable pairs. We hypothesized that targeted contrastive exposure to same-pitch adjacent-string pairs might improve generalization, and designed Comparison Training to test this. Section 6.2 reports a large matched-evaluation effect, after an initial confounded comparison had suggested otherwise.

### *5.2 Methodology*

Comparison Training adds dedicated comparison sessions to standard per-string recording. In each session, two strings sharing a pitch at a specific fret are played in deliberate alternation. The goal is to give the model balanced, temporally interleaved examples of the most confusable note pairs. The matched held-out evaluation in Section 6.2 measures the effect of including the comparison-session data, although the present design cannot isolate deliberate alternation from other differences in session composition.

We target the five canonical adjacent-string ambiguous pairs in standard EADGBE tuning:

- Open A2 (110 Hz) vs. E2 fret 5 (110 Hz)
- Open D3 (147 Hz) vs. A2 fret 5 (147 Hz)
- Open G3 (196 Hz) vs. D3 fret 5 (196 Hz)
- Open B3 (247 Hz) vs. G3 fret 4 (247 Hz)
- Open E4 (330 Hz) vs. B3 fret 5 (330 Hz)

These are the canonical low-fret adjacent-string ambiguities in standard tuning. Higher-fret ambiguous pairs are left for future work.

### *5.3 Data Collection*

Three sessions were recorded using the Boss Katana Gen 3 clean preset via USB-C at 44.1 kHz:

- strings_clean_session1: 60,294 frames, full neck coverage across all six strings
- strings_open_session1: open strings, varied attack styles and dynamics
- strings_comparison_session1: all five adjacent pair comparison sessions

Silence gate discards frames with mean FFT magnitude at or below 2.0, applied at collection time in the browser recording interface. The dataset was class-balanced to approximately 16.7% per string class across 322,215 total frames. In the shuffled-split condition, data is shuffled before the 80/20 train-validation split; Section 6.3 shows this shuffling does not eliminate temporal leakage between adjacent frames of the same sustained note, and reports a recording-session-held-out evaluation as a substantially more informative alternative.

## 6. Results

Unless otherwise stated, the reported accuracies in this section are calculated from the raw six-class neural-network output before pitch-based candidate masking, confidence penalties, and temporal stability gating. The latency measurement in Section 3.3 covers 26-feature extraction and dense-network inference. The accuracy figures below evaluate raw six-class network predictions and do not include pitch-based candidate masking, confidence penalties, or temporal stability gating.

### *6.1 Ablation Study*

Two directly comparable feature conditions were evaluated across five random seeds (42, 0, 1, 2, 3), with tf.keras.utils.set_random_seed() controlling both weight initialization and data shuffling:

- Condition A: 13 features, all data (322,215 frames) — removes MFCCs while holding the dataset constant.
- Condition C: 26 features, all data (322,215 frames) — the full model.

A third condition using 26 features without comparison-session data was also run during the original analysis. Because it uses a different underlying dataset and therefore a different shuffled validation set, it is omitted from the matched MFCC ablation and discussed only as part of the corrected Comparison Training analysis (Section 6.2).

Within each seed, Conditions A and C share identical shuffled train-validation assignments, preserving a paired comparison; the split varies across seeds along with weight initialization, so the reported standard deviations reflect total run-to-run variation rather than initialization noise alone.

| Condition | E2 | A2 | D3 | G3 | B3 | E4 | Overall |
|---|---|---|---|---|---|---|---|
| A: 13-feat, all data | 95.27±0.69 | 87.84±1.74 | 90.54±1.89 | 89.28±0.58 | 94.83±0.61 | 95.97±0.63 | 92.09±0.50 |
| C: 26-feat, full model | 97.95±0.43 | 95.83±0.39 | 95.87±0.87 | 97.31±0.53 | 97.98±0.37 | 98.90±0.19 | 97.25±0.32 |

*Table 2. Per-string shuffled frame-level validation accuracy, mean ± std over 5 seeds (42, 0, 1, 2, 3). Within each seed, Conditions A and C are trained on the same underlying frames and share identical shuffled train-validation assignments, preserving a paired comparison between them. A third condition using 26 features without comparison-session data was also run, but because it uses a different dataset and therefore a different shuffled validation set, it is omitted from Table 2 and is not used to assess Comparison Training. The omitted Condition B and Condition C results were evaluated on different shuffled validation sets and therefore are not a valid basis for assessing Comparison Training. Section 6.2 reports a matched held-out evaluation of Comparison Training that does not rely on this omitted comparison.*

MFCCs remain the primary accuracy driver: removing them drops overall accuracy from 97.25 ± 0.32% to 92.09 ± 0.50%, a roughly five-point difference substantially larger than the observed run-to-run variation. The omitted Condition B and Condition C results used non-comparable shuffled validation sets and therefore are not used to assess Comparison Training. Section 6.2 reports the matched held-out analysis instead.

### *6.2 Comparison Training: Matched Held-Out Evaluation*

An earlier Condition B versus C comparison, now omitted from Table 2, evaluated each condition on a different independently shuffled validation set: Condition B's validation set contains roughly 7,200 frames per class, while Condition C's contains roughly 11,400. This is not a valid basis for measuring Comparison Training's effect. To eliminate this confound, we instead trained two models on strictly comparable terms and evaluated both on an identical recording session held out from training, early stopping, and model selection (strings_open_session1.json, never used in training by either model): a baseline model trained on strings_clean_session1.json alone (60,294 frames), and a comparison model trained on strings_clean_session1.json plus strings_comparison_session1.json (165,857 frames). Both were run across the same five seeds. For all matched held-out models, early stopping used an internal validation subset drawn exclusively from the corresponding training data; strings_open_session1.json was not accessed during training, early stopping, hyperparameter selection, or model selection, and was evaluated once per trained model, after training and model selection were complete.

| Condition | E2 | A2 | D3 | G3 | B3 | E4 | Overall |
|---|---|---|---|---|---|---|---|
| Baseline (clean only) | 97.55±0.53 | 58.50±4.24 | 51.43±7.83 | 40.12±6.26 | 43.55±4.61 | 73.30±2.93 | 60.76±1.19 |
| Comparison (clean+comparison) | 97.82±0.43 | 76.37±3.37 | 80.15±1.82 | 90.12±2.43 | 83.58±3.01 | 91.18±2.38 | 86.53±1.23 |

*Table 3. Baseline vs. Comparison Training, both evaluated on the identical held-out strings_open_session1.json, mean ± std over 5 seeds. Early stopping for both conditions used an internal validation subset drawn exclusively from the corresponding training data; strings_open_session1.json was not accessed during training, early stopping, or model selection, and was evaluated once per trained model, after training and model selection were complete.*

The paired per-seed overall delta (comparison minus baseline) is +25.78 ± 1.46 percentage points, positive in all five seeds -- a large, consistent effect that supersedes the earlier confounded comparison. However, the comparison condition also trains on 2.75x more data (165,857 vs. 60,294 frames), so this result alone cannot distinguish an effect of the alternating-pair structure from a simple effect of training-set size.

To separate these explanations, we ran a size-matched control: the comparison condition's training data was downsampled, stratified by string label, to exactly 60,294 frames -- matching the baseline's size -- while still drawing proportionally from both the clean and comparison sessions. Both conditions were retrained and re-evaluated on the same held-out strings_open_session1.json across the same five seeds.

| Condition | E2 | A2 | D3 | G3 | B3 | E4 | Overall |
|---|---|---|---|---|---|---|---|
| Baseline (clean only) | 97.55±0.53 | 58.50±4.24 | 51.43±7.83 | 40.12±6.26 | 43.55±4.61 | 73.30±2.93 | 60.76±1.19 |

| Size-matched comparison | 97.15±1.06 | 72.82±9.64 | 78.02±4.27 | 89.30±3.98 | 85.36±1.98 | 93.44±0.77 | 86.02±1.75 |
|---|---|---|---|---|---|---|---|

*Table 4. Size-matched control: both conditions trained on exactly 60,294 frames, evaluated on identical held-out data, mean ± std over 5 seeds.*

The size-matched paired overall delta is +25.26 ± 1.51 percentage points -- numerically similar to the unmatched result (+25.78 ± 1.46) despite eliminating the entire 2.75x data-volume difference. This shows that total training-set size alone does not explain the effect. It does not yet establish that deliberate pair alternation itself is the causal mechanism: the comparison session also broadens the range of playing conditions the model is trained on, including examples that may resemble aspects of the open-string evaluation session, so the improvement is more precisely described as associated with the inclusion of comparison-session data than as proof that alternation per se is responsible. Isolating the effect of alternation from the effect of broader session coverage would require a further control -- for example, an equal-sized session of additional non-alternating single-string recordings -- which we leave for future work. This is nonetheless the paper's most rigorously controlled empirical result to date.

Per-string deltas remain noisy at this sample size: A2 shows a mean gain of 17.87 ± 5.04 points (unmatched) and 14.32 ± 11.77 points (size-matched); D3 shows a larger and more variable gain (28.72 ± 9.23 unmatched, 26.59 ± 4.86 size-matched). Given standard deviations that are a substantial fraction of the means for several of these figures, per-string causal claims about which specific string benefits most, or by how much, are not well supported at five seeds. The overall effect, by contrast, is positive for all five seeds in each experiment and should be treated as the paper's primary finding regarding Comparison Training: including the comparison-session data materially improves performance on the held-out open-string recording session, through a mechanism not explained by training-set size alone, though the precise per-string allocation of that benefit remains an open question.

### *6.3 Recording-Session-Held-Out Evaluation*

The shuffled frame-level split used in Section 6.1 allows near-duplicate frames from the same sustained note to appear in both training and validation, since adjacent frames within a single held note differ only slightly. To evaluate performance when an entire recording session is withheld, we additionally trained on two full recording sessions (strings_clean_session1 and strings_comparison_session1) and evaluated on a third session held out from training and model selection (strings_open_session1), with no shuffling across the session boundary. This was run across the same five seeds. Because the held-out session emphasizes open strings and varied attack styles, this evaluation changes both the recording session and aspects of the performance protocol. It therefore measures robustness to a combined session-and-protocol shift rather than isolating temporal leakage as the sole variable.

| Split | E2 | A2 | D3 | G3 | B3 | E4 | Overall |
|---|---|---|---|---|---|---|---|
| Shuffled (Sec. 6.1, Cond. C) | 97.95±0.43 | 95.83±0.39 | 95.87±0.87 | 97.31±0.53 | 97.98±0.37 | 98.90±0.19 | 97.25±0.32 |
| Recording-session -held-out | 97.82±0.43 | 76.37±3.37 | 80.15±1.82 | 90.12±2.43 | 83.58±3.01 | 91.18±2.38 | 86.53±1.23 |

*Table 5. Shuffled frame-level split vs. recording-session-held-out evaluation, mean ± std over 5 seeds. The recording-session-held-out condition trains on strings_clean_session1 and strings_comparison_session1, and evaluates on strings_open_session1, held out from training and model selection.*

The recording-session-held-out evaluation yields 86.53 ± 1.23% overall, a 10.72-point gap that is substantially larger than the observed run-to-run variation across the five seeds. Early stopping for this experiment used an internal validation subset drawn exclusively from the training data; strings_open_session1.json was evaluated once, after training and model selection were complete. A2 and D3 show the two largest accuracy drops under this evaluation. Their reduced accuracies are consistent with the difficulty of these classes observed in Section 6.2, although only A2 exhibits one of the largest standard deviations across seeds. This result shows that performance decreases substantially when an entire recording session is withheld. The decrease is consistent with adjacent-frame leakage inflating the shuffled result, although session-specific and protocol-specific distribution shifts may also contribute.

### *6.4 Held-Out Free-Play Evaluation*

A separate held-out session of approximately 15 minutes of free-play guitar was recorded after all training was complete — same guitar, same player, same clean preset, never seen during training or hyperparameter selection. Ground-truth string labels were recorded through the browser toggle interface and aligned with the corresponding captured frames; frames were assigned the label active at the time of capture, with no exclusion window applied around toggle transitions. Evaluation retained only frames satisfying the same silence-gate criterion used during training (mean FFT magnitude greater than 2.0); no additional pitch-validity filtering was applied. The resulting class distribution was not explicitly balanced, as the free-play session was not designed for even string coverage. Reported accuracy uses raw six-class classifier output, consistent with the other figures in this section (see note at the start of Section 6). The held-out set contains approximately 103,000 frames.

| Evaluation | E2 | A2 | D3 | G3 | B3 | E4 | Overall |
|---|---|---|---|---|---|---|---|
| Shuffled validation (Sec. 6.1) | 97.95±0.43 | 95.83±0.39 | 95.87±0.87 | 97.31±0.53 | 97.98±0.37 | 98.90±0.19 | 97.25±0.32 |
| Recording-session -held-out (Sec. 6.3) | 97.82±0.43 | 76.37±3.37 | 80.15±1.82 | 90.12±2.43 | 83.58±3.01 | 91.18±2.38 | 86.53±1.23 |
| Held-out free-play | 95.7% | 82.3% | 90.5% | 90.7% | 91.7% | 76.3% | 87.8% |

*Table 6. Three evaluation conditions using the same architecture and feature family. The shuffled and recording-session-held-out figures are five-seed results from the revised experimental configuration (Sections 6.1,*

*6.3), while the held-out free-play result evaluates the originally deployed model (Section 6.5), trained under a different early-stopping configuration. The recording-session-held-out and held-out free-play results are numerically close (86.53% vs. 87.8%) despite evaluating different trained weights and differing in what makes the data unseen. This proximity is encouraging and suggests that ordinary session-to-session variation may account for a substantial portion of the gap from shuffled validation. However, one held-out session and one free-play recording, on two different trained models, are insufficient to separate session effects from differences in playing protocol conclusively.*

The unweighted macro-average of the six per-string accuracies is 87.9%, closely matching the frame-weighted overall accuracy of 87.8%, indicating the result is not substantially driven by class imbalance in the unbalanced free-play session. The two largest held-out free-play failures are E4→B3 (15.7% error rate) and A2→E2 (13.9% error rate), both adjacent-string pairs. Because prior work reports event-level F-measure across multiple instruments and players under different evaluation protocols, none of the three figures in Table 6 are directly comparable to the F = 0.90 benchmark; each is an internal diagnostic result for this single-instrument prototype, and the session-held-out and free-play figures are the more defensible generalization estimates for this prototype.

### *6.5 Model Architecture*

The final model: Input(26) → Dense(128, ReLU) → Dropout(0.3) → Dense(32, ReLU) → Dropout(0.2) → Dense(6, softmax). Training uses the Adam optimizer, batch size 32, and early stopping monitoring validation accuracy (not validation loss) with restore_best_weights=True. Two configurations were used across the reported experiments: the originally deployed model used patience=5 with a maximum of 25 epochs, while the revised ablation, matched Comparison Training, size-matched control, and recording-session-held-out experiments in this revision use patience=6 with a maximum of 40 epochs. The size-matched subset was resampled independently for each seed using that seed's deterministic random state. Class weights are balanced via sklearn compute_class_weight.

A preliminary 1D CNN baseline was explored but is not reported quantitatively here; systematic comparison with learned spectral models is left for future work. FFT output is already a feature vector, so convolutional translation-invariance along the frequency axis is not obviously well-motivated for this input, which informed the choice to proceed with the dense architecture over engineered features.

### *6.6 Comparison to Prior Work*

| System | Classifier | Metric | Score |
|---|---|---|---|
| Abeßer (2013; CMMR 2012) | SVM + spectral envelope | F-measure | 0.90 |
| Geib et al. (2017) — SIF | FFT/SHS + SIF features | F1 | 0.72 |
| Kehling et al. (2014) | Tablature pipeline (string est.) | String Acc. | 82% |
| Fretiq — shuffled split (this work) | Dense NN + MFCC features | Frame-level Val. Acc. | 97.25±0.32% |

| Fretiq — session held-out (this work) | Dense NN + MFCC features | Session-held-out Acc. | 86.53±1.23% |
|---|---|---|---|
| Fretiq — held-out (this work) | Dense NN + MFCC features | Held-out Free-Play Acc. | 87.8% |

*Table 7. Comparison with prior work. Prior works report event-level F-measure or string accuracy across multiple instruments. Fretiq reports frame-level shuffled-validation or recording-session-held-out accuracy on a single instrument and player. Metrics are not directly comparable.*

## 7. Implementation: Critical Failure Modes

Two critical failures during development are documented here as reproducibility contributions — both are likely to recur in browser-based ML inference systems built on similar stacks.

### *7.1 Training-Inference Feature Dimensionality Mismatch*

The initial TypeScript implementation computed only 13 features while Python training computed all 26. The model received a 26-dimensional vector with the final 13 dimensions always zero. Because the dense weights still produced outputs, nothing obviously broke — the system appeared to run while predicting on garbage input.

The failure surfaced through softmax logging: confidence was uniformly distributed across strings regardless of input. Fixing it required implementing the full mel filterbank and DCT-II MFCC pipeline in TypeScript and verifying feature vectors frame-by-frame against the Python output. The ablation study retrospectively quantifies the cost: 92.09±0.50% overall accuracy at 13 features versus 97.25±0.32% at 26.

### *7.2 Model Serialization Incompatibility*

TensorFlow.js's tf.loadLayersModel() could not parse the exported model.json — written using Keras 2's model.to_json() format — and raised an internal deserialization error without a clear root cause.

The fix bypasses the JSON parser entirely: fetch weights.bin as an ArrayBuffer, reconstruct the architecture manually via tf.sequential(), and call model.setWeights() with Float32Array tensors sliced at the correct byte offsets. The full weight layout — kernel and bias for three dense layers — totals 7,782 float32 values, or 31,128 bytes.

## 8. Limitations and Future Work

Several limitations bound what these results can claim.

Single instrument, player, and preset. Every frame in this study — training, validation, and held-out — comes from one guitar, one player, and one amplifier setting. Timbre varies substantially across body styles, pickup configurations, string gauges, and attack characteristics. All accuracy figures should be read in this narrow context.

Shuffled-split optimism under session holdout. Section 6.3 shows the shuffled split produces an estimate roughly 10.7 percentage points higher than the recording-session-held-out result, and that this recording-session-held-out figure is numerically close to an independently collected held-out free-play result. The shuffled 97.25±0.32% figure should not be interpreted as a generalization estimate; the session-held-out or free-play figures are the more appropriate estimates of performance under unseen recording conditions.

Preset generalization failure. On the Boss Katana distortion preset without retraining, the model produced high-confidence wrong predictions across all positions — not low-confidence uncertainty, but confident domain shift. Distortion clipping collapses the spectral and MFCC structure the model relies on. Retraining on combined presets is the planned fix.

Session-to-session vs. free-play-specific generalization gap. The numerical closeness between the recording-session-held-out evaluation (86.53±1.23%) and held-out free-play (87.8%) results is encouraging and suggests that ordinary session-to-session variation may account for a substantial portion of the gap from shuffled validation. However, one held-out session and one free-play recording are insufficient to separate session effects from differences in playing protocol conclusively. Temporal models that integrate evidence across consecutive frames remain a promising direction. Evaluating true transition modeling will require new recordings containing natural string switches, slides, hammer-ons, pull-offs, and other transitions, because the current training sessions consist primarily of sustained single-string runs.

Comparison Training: matched-evaluation effect and remaining open questions. A matched-evaluation experiment (Section 6.2) finds that including comparison-session data improves held-out-session accuracy by roughly 26 percentage points, and a size-matched control shows that total training-set size alone does not explain the effect. This is a substantially better-controlled result than the paper's original ablation, and it supersedes the original conclusion of no net benefit, which followed from comparing training conditions on different, non-comparable shuffled validation sets. What remains open is whether the deliberate alternation of adjacent-string pairs specifically drives the improvement, or whether differences in pitch and fret-position coverage, attack or dynamic variation, or other session-specific characteristics of the comparison-session data are doing some or all of the work; the present design cannot fully separate these explanations. The precise per-string allocation of the effect is also unresolved: per-seed deltas for individual strings (A2, D3) are noisy enough, with occasional sign flips, that claims about which specific strings benefit most are not well supported at five seeds. Systematic sessions covering all five targeted adjacent-string pairs, an additional non-alternating control session of matched size, and more seeds, are needed to resolve these questions.

Reproducibility. The training pipeline, feature extraction code, evaluation scripts, and model weights are publicly available at https://github.com/D1gits0/fretiq, with full environment documentation in REPRODUCE.md. Multi-seed reproduction uses

tf.keras.utils.set_random_seed(), which was required to obtain bit-identical repeated runs; tf.random.set_seed() alone was found insufficient for Keras 2.13's initializer RNG.

## 9. Conclusion

Fretiq is a preliminary single-instrument, single-player browser-native electric guitar string classification system. Across five seeds, it achieves 97.25±0.32% shuffled frame-level validation accuracy and 86.53 ± 1.23% recording-session-held-out accuracy, the latter numerically close to an independently collected 87.8% held-out free-play result. It runs entirely in-browser via USB-C direct input — no hexaphonic pickup, fretboard sensor, camera, or multi-microphone setup required.

MFCCs are the primary accuracy driver, accounting for roughly a 5-point gain over a 13-feature baseline that is far larger than the observed run-to-run variation across five seeds. An initial Comparison Training comparison using non-comparable shuffled validation sets suggested no net benefit; a matched-evaluation reanalysis supersedes this, finding a large, consistent improvement of roughly 26 percentage points on an unseen recording session, positive for all five seeds in both the unmatched and size-matched experiments. This shows the effect is not explained by training-set size alone, though it does not yet establish that the deliberate alternation of adjacent-string pairs, rather than differences in pitch and fret-position coverage, attack or dynamic variation, or other session-specific characteristics, is the specific mechanism. The precise per-string allocation of this benefit also remains uncertain at five seeds. The close numerical agreement between the recording-session-held-out evaluation and held-out free-play accuracy is encouraging and suggests that ordinary session-to-session variation may account for a substantial portion of the roughly 10.7-point generalization gap, though a single held-out session and a single free-play recording are insufficient to separate session effects from playing-protocol differences conclusively. This motivates the temporal modeling discussion in Section 8.

Because prior work reports event-level F-measure across multiple instruments and players, the accuracy figures here are not directly comparable to published benchmarks — they are internal diagnostic results for a single-instrument prototype. Two runtime failure modes are documented in detail. The system demonstrates that real-time string-level guitar classification is achievable in a standard web browser from a single audio stream, and this revision identifies, with quantified uncertainty, the conditions under which its reported accuracy is and is not representative of real deployment performance.

### *Revision note for arXiv v2*

This revision replaces previously reported single-run ablation results with five-seed averages, corrects the description of the early-stopping criterion, and adds a recording-session-held-out evaluation. An initial version of that recording-session-held-out evaluation and the initial Comparison Training experiments inadvertently used strings_open_session1.json as the

validation_data monitored by early stopping, meaning model selection was informed by the nominally held-out session; all affected reported experiments (session-held-out dense, unmatched Comparison Training, and the size-matched control) were rerun with early stopping restricted to an internal split of the training data only, with strings_open_session1.json evaluated once per model after training and model selection were complete. An exploratory LSTM analysis, not retained in the revised manuscript, was corrected separately. The corrected numbers reported here are the rerun figures; in this case correcting the leakage strengthened rather than weakened the paper's central finding. This revision also supersedes the paper's original conclusion regarding Comparison Training: the original comparison evaluated training conditions on different, non-comparable shuffled validation sets and found no aggregate benefit; the corrected matched-session evaluation supersedes that analysis and shows that including comparison-session data improves performance substantially on the common open-string evaluation session, positive for all five seeds in each of the unmatched and size-matched experiments. Equalizing training-set size preserved the gain, so frame count alone cannot explain it, although the study does not isolate deliberate alternation from other differences in session composition, such as pitch and fret-position coverage or attack and dynamic variation. The central system architecture and held-out free-play evaluation are unchanged. We consider these corrections a significant improvement in the reliability of the paper's central methodological claim, made possible by a reader identifying both the original evaluation-set mismatch and the subsequent early-stopping leakage.